\documentclass[aps,preprint]{revtex4}

\usepackage{graphicx}

\newcommand{\be}{\begin{equation}}
\newcommand{\ee}{\end{equation}}

\begin{document}

\title{
\bf Critical Behavior of Liquid $^3He$ }
\author{{\bf G.H. Bordbar,   S.M. Zebarjad  and F. Shojaei}
}
 \affiliation{ Department of Physics, Shiraz University,
Shiraz 71454, Iran\\
and\\
Institute for Studies in Theoretical Physics and Mathematics (IPM),\\
Tehran, P. O. Box 19395-5531, Iran}

%************************************************************************
\begin{abstract}
We  investigate the liquid-gas  second-order phase transition in
liquid $^3He$  using the variational calculations based on the
cluster expansion of the energy functional. We  also  compute the
critical point exponents of liquid $^3He$ which are in agreement
with experimental data.
\end{abstract}
\maketitle
%***********************************************************************

\section{Introduction} \label{intro} The liquids $^3He$ and
$^4He$ are the only quantum liquids which exist naturally. The
word ``quantum liquid" comes from the fact that for these systems,
the interatomic distance is at the order of their De Broglie
wavelenght. $^3He$ is a liquid of strongly interacting fermionic
atoms which behaves quite differently than the normal liquids  at
low temperature (Wilks, 1970; Kent, 1993). The properties of
liquid $^3He$ have been studied  using different many-body
techniques (Clark and Westhaus, 1966; Nafari and Doroudi, 1995;
Luijen and Meyer, 2000; Kindermann and Wetterich, 2001; Takano and
Yamada, 1994; Viviani et al., 1988; Pricaupenko and Treiner, 1995;
Fantoni et al., 1982; Krotscheck and Smith, 1983; Friman and
Krotscheck, 1982). Recently, the behavior of liquid $^3He$ near
its critical point has been investigated using path-integral
molecular dynamics and quantum virial expansion (M\"{u}er and
Luijten, 2002).

One of the most powerful techniques in many-body calculations is
the variational method which is based on the cluster expansion of
the energy functional (lowest order constrained  variational
method) (Owen et al., 1977; Bordbar and Modarres, 1997, 1998;
Modarres and Bordbar, 1998; Bordbar and Riazi, 2001, 2002;
Bordbar, 2002a, 2002b, 2003, 2004; Bordbar and Hashemi, 2002).
This is a fully self-consistent method and does not introduce any
free parameter to the calculations. The crucial point in this
method is the functional minimization with respect to the two-body
correlation function subjected to the normalization constraint
which finally leads to a Euler-Lagrange differential equation. The
convergence of its results has been shown by computing the
three-body  cluster energy term (Bordbar and Modarres, 1997).

The liquid-gas phase transition near the critical point
(second-order phase transition) is an interesting subject in
statistical mechanics. This behavior, critical phenomena, is
caused by the existence of singularity in thermodynamic functions
of the system at the transition point. The nature of these
singularities in various  measurable quantities at the critical
point is described by  the critical exponents.

In our previous paper, we have calculated some thermodynamic
properties of liquid  $^3He$ using the  variational  method shows
a nice agreement between experimental data and calculated results
especially for free energy and entropy (Bordbar and Hashemi,
2002).
 In this article, we present the
critical behavior of the liquid $^3He$. We organize the paper as
follows: In section \ref{Crit}, we obtain the critical properties
of liquid $^3He$ by calculating the critical isotherm. We
investigate  the critical behavior of liquid $^3He$ by computing
the critical point exponents in section \ref{exponent}.
%**********************************************************************
\section{Critical Isothermal Equation of State}
\label{Crit}
\begin{figure}
\includegraphics[height=13cm]{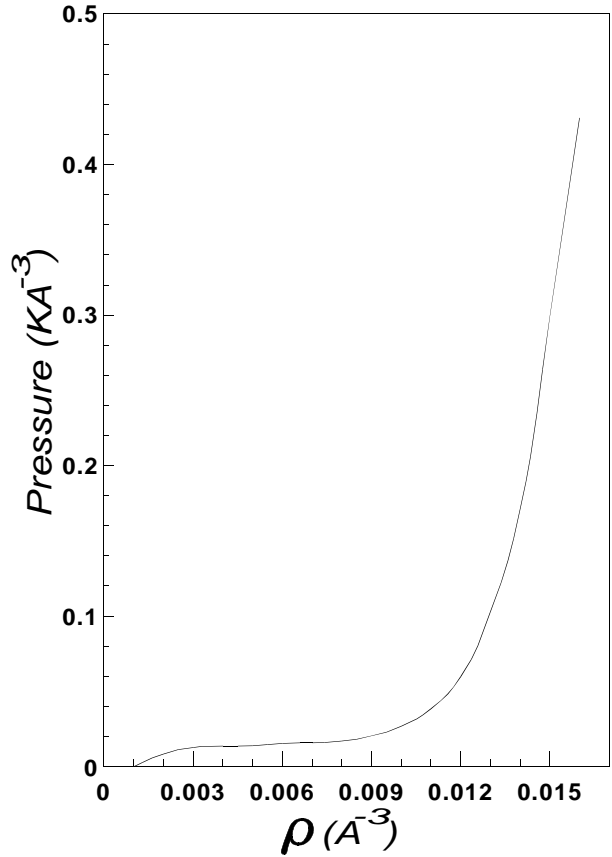}
 \caption{The
critical equation of state for liquid $^3He$.}
 \label{pressure}
\end{figure}

 The equation of state is the key point for
investigating the second-order phase transition in a hydrostatic
system. The isothermal equation of state can be calculated from
the Helmholtz free energy, $F$:
\begin{equation}
P=\rho^2\frac{\partial F}{\partial\rho}|_T,
\end{equation}
where $P$, $T$ and $\rho=\frac{N}{V}$ are the pressure,
temperature and number density, respectively. $N$ and $V$ are the
total number of particles and volume. To obtain the free energy of
liquid $^3He$, we use the variational method explained in
Appendix. The result for the equation of state at the critical
temperature (critical isotherm) is shown in Fig. (\ref{pressure}).
As seen from the Fig. (\ref{pressure}), at the critical point, the
isotherm curve shows an inflection point which satisfies:
\begin{equation}
\frac{\partial P}{\partial \rho}|_{T_c}=\frac{\partial^2
P}{\partial \rho^2}|_{T_c}=0,
\end{equation}
where $T_c$ is the critical temperature. The calculated  critical
temperature, density ($\rho_c$) and pressure ($P_c$) of liquid
$^3He$ are presented in Table \ref{Tab1}. The experimental results
(Heller, 1967; Fisher, 1967; Pittman et al., 1979) are also given
for comparison. We can see a good agreement between these results.
\begin{table}
\begin{center}
\caption{Critical point properties of liquid $^3He$. }
\label{Tab1}
\begin{tabular}{|c|c|c|c|}
  \hline
   & $T_c(K)$ &  $\rho_c(A^{-3})$& $P_c(KA^{-3})$ \\
  \hline
  Our results & 4.36 & 0.0054 & 0.0139 \\
  Exp. results (Heller, 1967) & 3.324 & 0.00834 &  0.00844 \\
  Exp. results (Pittman et al., 1979)&3.317&0.00827&0.00846\\
  \hline
\end{tabular}
\end{center}
\end{table}
%***********************************************************
\section{Critical Exponents}
\label{exponent} 
\begin{figure}
\includegraphics[height=12cm]{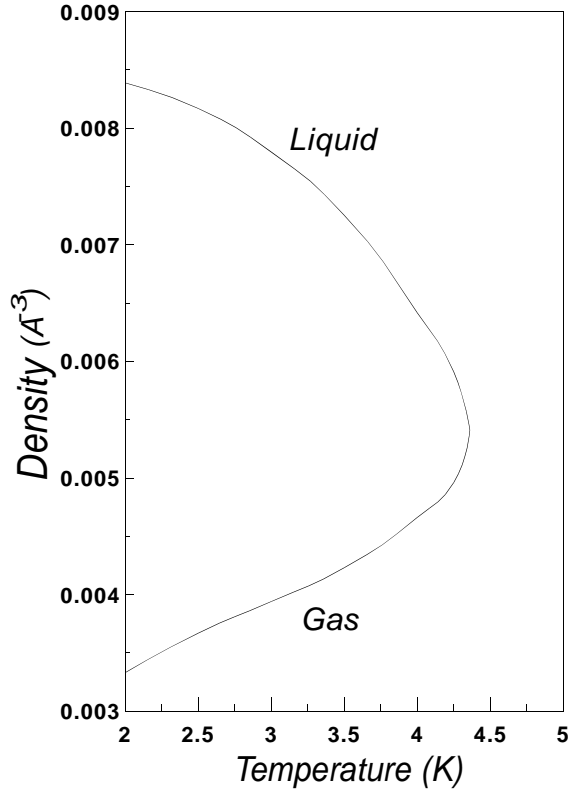}
 \caption{The
liquid and gas densities versus  temperature for $^3He$.}
\label{density}
\end{figure}

For a hydrostatic system, the two-phase
coexistence conditions are
 \begin{eqnarray}
   P_{liquid}&=&P_{gas}  \nonumber \\
  \mu_{liquid} &=& \mu_{gas},
\end{eqnarray}
where the $\mu_{liquid}$ and $\mu_{gas}$ are the chemical
potential of liquid and gas phases respectively. As the
temperature increases, the liquid density   decreases and the gas
density  increases. At the critical temperature these densities
become equal to each other. This behavior  for $^3He$ is shown in
Fig. (\ref{density}).
The order parameter $\rho_{liquid}-\rho_{gas}$  which is defined
to investigate the critical behavior of this system  vanishes at
the critical point. However other thermodynamic properties diverge
at this point. The critical point exponents are defined to  study
the asymptotic behavior of singular thermodynamic functions near
the critical point. For this purpose,  the following functions for
the thermodynamic quantities are introduced (Garrod, 1995):
 \begin{itemize}
     \item \textbf{Order parameter}\\We can define the exponent
           $\beta$ for this parameter as follows:
            \begin{equation}
            \rho_{liquid}-\rho_{gas}\sim (-\epsilon)^{\beta};
            \,\,\,\,\,\,\,\,\,\,\, \epsilon\longrightarrow 0^-,
            \end{equation}
            where
            \begin{equation}
            \epsilon=\frac{T-T_c}{T_c}.
            \end{equation}
          The critical exponent $\beta$  characterizes the behavior
          of the order parameter and of course, the above function
          is meaningful
          only below the critical point in the region where the
          order parameter  is not zero. To obtain $\beta$, we
          draw the order parameter as a function of $\epsilon$
          on the log-log scale in Fig. (\ref{beta}).
          \begin{figure}
           \includegraphics[height=12cm]{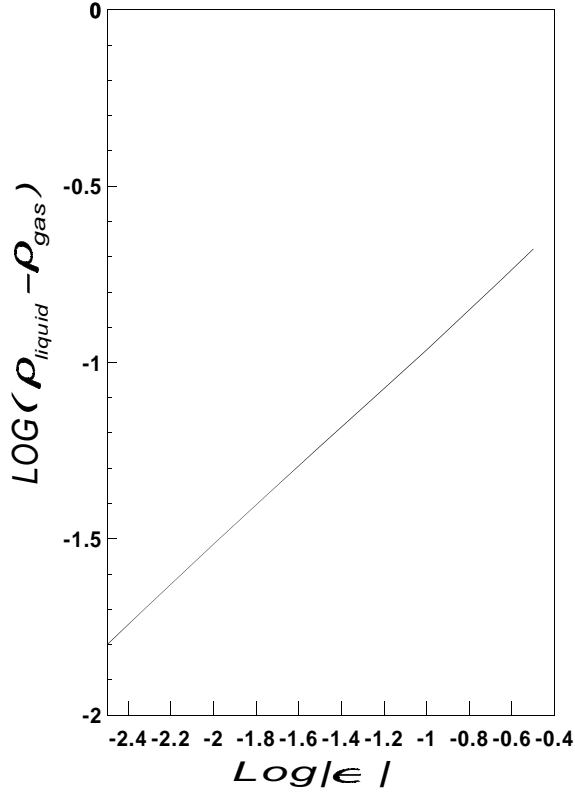}
           \caption{The
           order parameter versus $\epsilon$ on log-log scale for $^3He$.}
          \label{beta}
           \end{figure}
           The slope of this figure yields  the value of
           $\beta=0.56239\pm 0.01386$.
    \item \textbf{Pressure}\\
         By defining the exponent $\delta$, we can describe the
         critical isotherm
         \begin{equation}
            P-P_c\sim (\rho-\rho_c)^{\delta};
            \,\,\,\,\,\,\,\,\,\,\, \rho\longrightarrow \rho_c,
            \end{equation}
            where  $\epsilon=0$ ($T=T_c$). In Fig.
            (\ref{delta}), $P-P_c$ as a function of $\rho-\rho_c$ is
            shown.
           \begin{figure}
           \includegraphics[height=12cm]{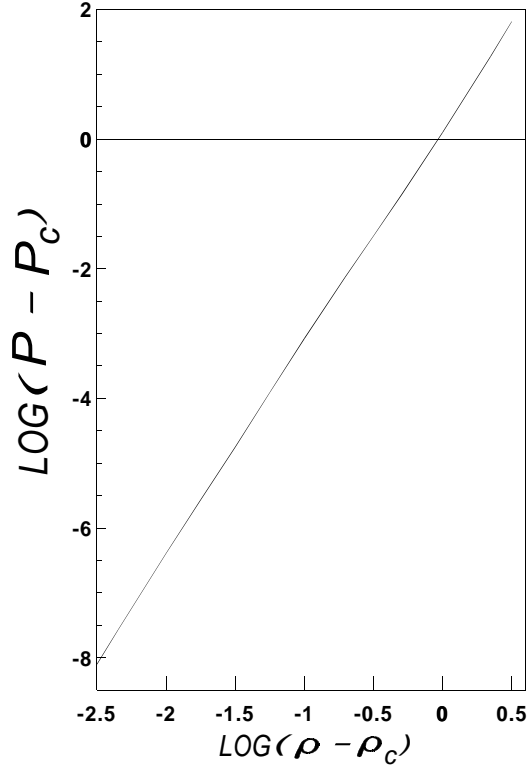}
           \caption{The $P-P_c$ versus  $\rho-\rho_c$
            at critical temperature ($T_c$) for $^3He$.}
          \label{delta}
           \end{figure}
           The value of $\delta$ obtained from this figure
           is $3.31032\pm 0.08192$.
    \item \textbf{Heat Capacity}\\
          The exponent $\alpha'$ and $\alpha$ characterize the behavior of
          specific heat ($C_V$) below and above the critical temperature
          respectively along the critical isochore ($V=V_c$)
         \begin{eqnarray}
         C_{V_c}&=&(-\epsilon)^{-\alpha'};\,\,\,\,\,\,\,\,\,\,\,
         \epsilon\longrightarrow 0^-,  \nonumber \\
          C_{V_c}&=&(\epsilon)^{-\alpha};\,\,\,\,\,\,\,\,\,\,\,
          \epsilon\longrightarrow 0^+.
         \end{eqnarray}
         In Fig. (\ref{alpha}), the specific heat  along the critical
         isochore versus $\epsilon$ is
         shown.
          \begin{figure}
           \includegraphics[height=12cm]{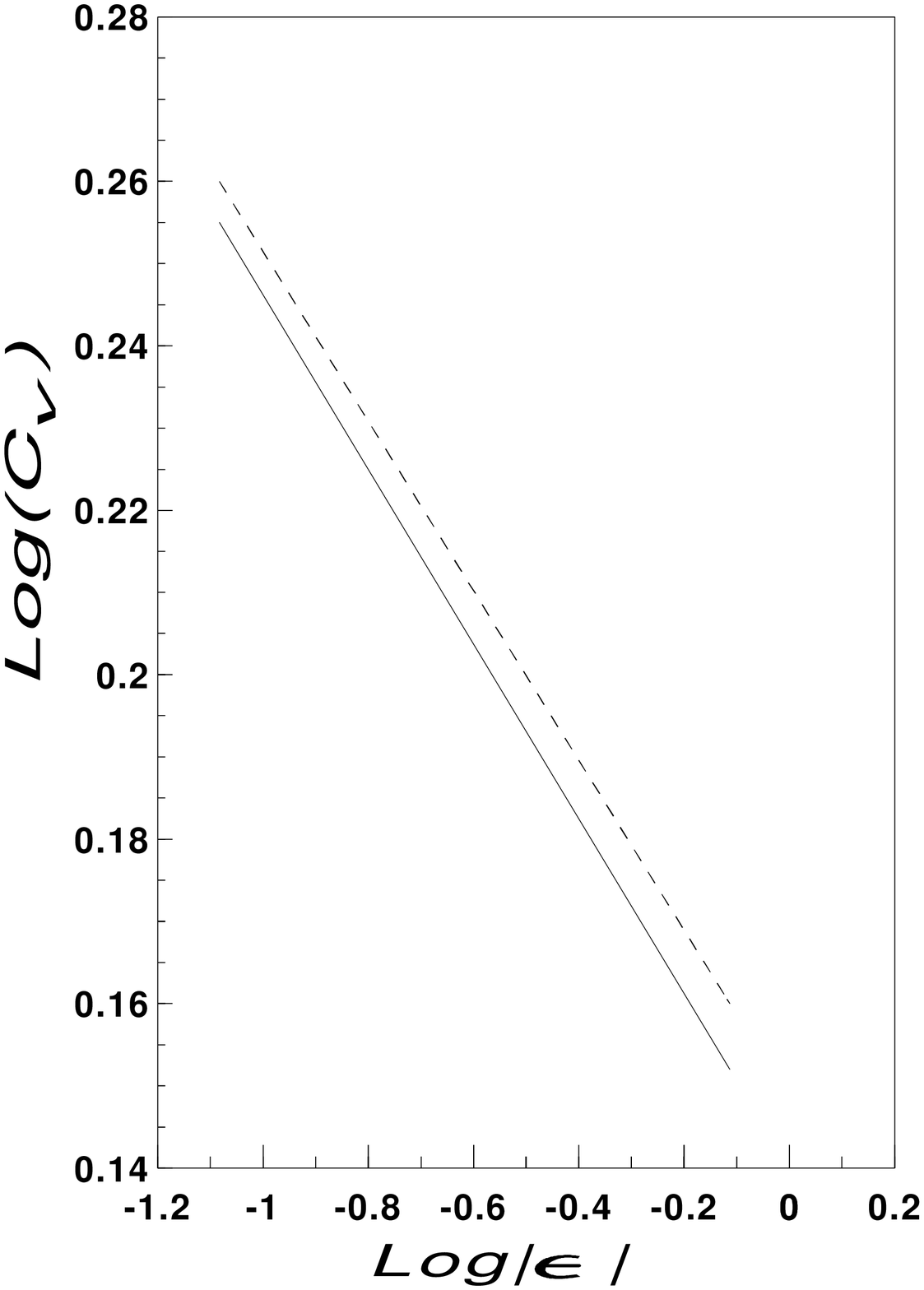}
           \caption{Specific heat along the critical isochore as a function of $\epsilon$
           above (full curve)  and below (dashed curve)
           critical temperature for $^3He$.}
          \label{alpha}
           \end{figure}
           The values $\alpha'=0.1018\pm 0.0001$ and $\alpha=0.10609\pm
           0.0014$ are extracted from the Fig. (\ref{alpha}).
    \item  \textbf{Isothermal Compressibility}\\
    For describing the behavior of isothermal compressibility ($K$) near
    the critical point, the exponents $\gamma$ and $\gamma'$ are
    defined to be
       \begin{eqnarray}
         K&=&(-\epsilon)^{-\gamma'}; \,\,\,\,\,\,\,\,\,\,\,
         \epsilon\longrightarrow 0^-,  \nonumber \\
          K&=&(\epsilon)^{-\gamma};\,\,\,\,\,\,\,\,\,\,\,
          \epsilon\longrightarrow 0^+.
         \end{eqnarray}
         The calculated values of isothermal compressibility shown
         in Fig. (\ref{gamma}) leads to $\gamma= 1.05343\pm
         0.01077$ and
         $\gamma'= 1.05343\pm 0.01077$.
          \begin{figure}
           \includegraphics[height=12cm]{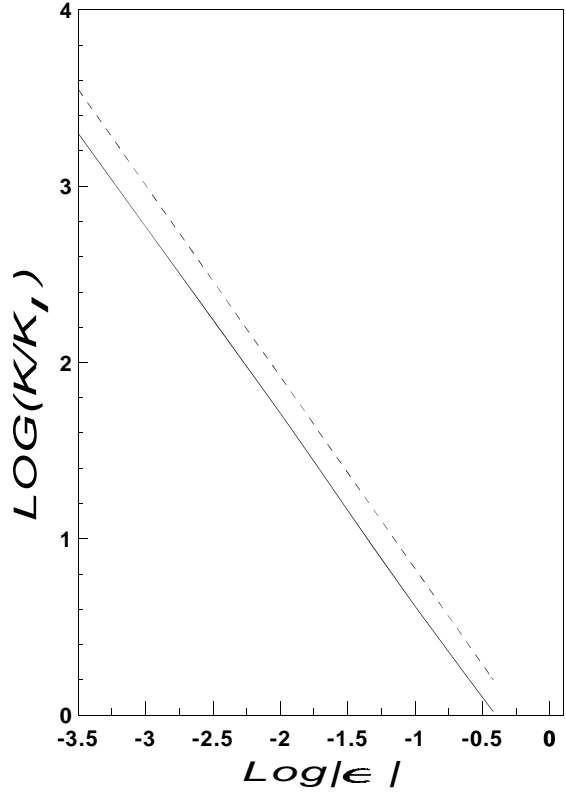}
           \caption{ Isothermal compressibility as a function of
           $\epsilon$ above (full curve)  and
           below (dashed curve)  critical temperature for $^3He$.
           $K_I$ is the ideal fermi gas compressibility at $\rho=\rho_c$
           and $T=T_c$.}
          \label{gamma}
           \end{figure}
\end{itemize}

We have presented the whole critical exponents for the $^3He $ in
Table \ref{exp}.
\begin{table}
\caption{Critical exponents for $^3He$. }
 \label{exp}
\begin{center}
{\small
\begin{tabular}{|c|c|c|c|c|c|c|}
  \hline
     & $\beta$ & $\delta$ & $\alpha'$ & $\alpha$ & $\gamma$ & $\gamma'$\\
  \hline
  Our results &$0.5624$ & $3.3103$ & $0.1018$ &$0.1061$  &
  $1.0534$  &$1.0560$\\
  &$\pm 0.0139$ & $\pm0.0819$ & $\pm0.0001$ &$\pm0.0014$  & $\pm 0.0108$
  &$\pm 0.0093$\\
\hline
  Exp. results (Heller, 1967) &$\sim 0.361$ & $\sim 4.21$ & $\sim 0.105$
  & $\sim 0.105$ & $\sim 1.17$ & $\sim 1.17$ \\
\hline
  Exp. results (Pittman et al., 1979)&$0.322\pm 0.002$&---&---&---&
  $1.19\pm 0.01$&---\\
     \hline
\end{tabular}
}
\end{center}
\end{table}
The experimental results (Heller, 1967; Fisher, 1967; Pittman et
al., 1979) are also given for the comparison in Table \ref{exp}.
There is a good agreement between our calculations for the
critical exponents and the experimental results. From Table
\ref{exp}, it can be seen that the Griffiths and Rushbrooke
inequalities (Huang, 1987; Griffiths, 1965) are satisfied by our
results for the critical exponents of $^3He$,
\begin{eqnarray}
  \alpha+2\beta+\gamma& \geq & 2  \nonumber\\
  \alpha+\beta(1+\delta) &\geq& 2.
\end{eqnarray}
%*******************************************************************
\section{Summary and Conclusion}
The liquid-gas phase transition near the critical point is of
special interest in statistical mechanics.  In this work, we have
 computed the critical equation of state for liquid $^3He$ which led to
 critical density, temperature and pressure of this system. The
 critical exponents, $\beta$, $\delta$, $\alpha$ and $\gamma$ for
 this system are computed.
The calculated critical exponents satisfies the Griffiths and
Rushbrooke inequalities.
  A comparison between our results and
 experimental data is made which  shows a good agreement between
 theoretical calculation and experimental results.
%**********************************************************************
\acknowledgements{ Financial support from Shiraz University
research council and IPM is gratefully acknowledged.}
%%%%%%%%%%%%%%%%%%%%%%%%%%%%%%%%%%%%%%%%%%%%%%%%%%%%%%%%%%%%%%%%%%%%%%%%%%%
\appendix
\section*{Appendix}
In this appendix, we give a  brief review to obtain the free
energy of liquid $^3He$  using  the lowest order  constrained
variational method based on the cluster expansion of the energy
functional (Owen et al., 1977; Bordbar and Modarres, 1997, 1998;
Modarres and Bordbar, 1998; Bordbar and Riazi, 2001, 2002;
Bordbar, 2002a, 2002b, 2003, 2004; Bordbar and Hashemi, 2002). In
this method, We choose a trial many-body wavefunction   as
\begin{equation}\label{psi}
    \Psi= \{\prod_{i< j } f(ij)\} \Phi,
\end{equation}
  where $f(ij)$ is the two-body correlation function and
$\Phi$ is the Slater determinant of noninteracting particles
wave-functions (plane waves). We then apply  the cluster expansion
to the energy per particle (Clark, 1979) and keep one and two-body
energy terms,
%\footnote{It is shown that higher order energy terms
%are ignorable when  we are dealing with short-range potential
%(Bordbar and Modarres, 1997).}
\begin{equation}
E=\frac{1}{N}\frac{\langle\Psi|H|\Psi\rangle}{\langle\Psi|\Psi\rangle}=E_1+E_2,
\end{equation}
where
\begin{eqnarray}\label{energy2}
    E_1&=&\sum_{i}\frac{\hbar^2k_i^2}{2m}n(k_i),\\ \label{energy3}
    E_2&=&\frac{1}{2N}\sum_{ij}<ij\mid w(12)\mid ij-ji>.
\end{eqnarray}
 In the above equations,  $n(k_i)$ is the Fermi-Dirac  distribution function
and
\begin{equation}
    w(12)= \frac{\hbar^2}{m}\bigg(\nabla_{12}
f(12)\bigg)^2+f^2(12) V(12),
\end{equation}
where $V(12)$  is the interatomic potential. In the thermodynamic
limit, Eqs. (\ref{energy2}) and (\ref{energy3}) read:
 \begin{eqnarray} \label{energy4}
     E_1 &=& \frac{\hbar^2}{2m\rho\pi^2}\int_0^\infty n(k)k^4dk,\\
     \label{energy5}
     E_2 &=& \frac{2\pi
     \rho\hbar^2}{m}\int_0^\infty\bigg[f'^2(r)+\frac{m}{\hbar^2}f^2(r)V(r)\bigg]
     \bigg[1-\frac{1}{2}\bigg(\frac{\gamma(r)}{\rho}\bigg)^2\bigg]r^2dr,
 \end{eqnarray}
 where $\rho$ is the number density,
\begin{equation}
\rho=\frac{1}{\pi^2}\int_0^\infty n(k)k^2dk,
\end{equation}
\begin{equation}
    \gamma(r)=\frac{1}{\pi^2}\int_0^\infty \frac{\sin
kr}{kr}n(k)k^2dk,
\end{equation}
 and $f'(r)=\frac{\partial f(r)}{\partial r}$.

 At this point, we  minimize the energy
functional with respect to the two-body correlation function,
$f(r)$, to obtain  the following Euler-Lagrange differential
equation
\begin{equation}\label{pd}
    \bigg\{\frac{2m}{\hbar^2}f(r)V(r)
    +2\lambda f(r)\bigg\}\bigg[1-\frac{1}{2}\bigg(\frac{\gamma(r)}{\rho}\bigg)^2\bigg]
    -\frac{\partial}{\partial r}\bigg[ 2f'(r)\bigg[1-\frac{1}{2}\bigg(\frac{\gamma(r)}
    {\rho}\bigg)^2\bigg] \bigg]=0,
\end{equation}
where $\lambda$ is the Lagrange multiplier which imposed the
normalization condition $\langle\Psi|\Psi\rangle=1$. By solving
Eq. (\ref{pd}), using the numerical technique, the two-body
correlation function, $f(r)$ and therefore the energy of the
system are obtained. This  finally leads to the free energy
function of the system
\begin{equation}
F=E-TS,
\end{equation}
where $T$ and $S$ are the temperature and entropy per particle of
the systems (Fetter and Walecka, 1971).

\begin{figure}
\includegraphics[height=15cm]{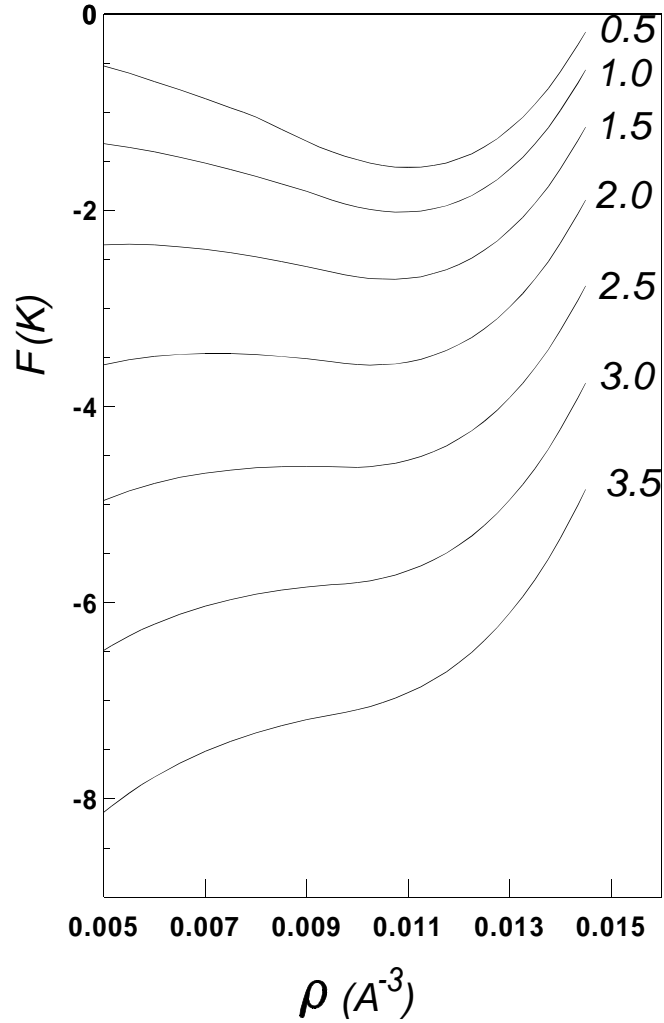}
\caption{The free energy of liquid $^3He$ as a function of number
density at different temperatures.}
\label{free}
\end{figure}

To  calculate the free energy of liquid $^3He$, we use the Aziz
interatomic potential (Aziz et al., 1979) in Eqs. (\ref{energy5})
and (\ref{pd})
 \begin{equation}
V(r) = \epsilon\left\{ Ae^{-\alpha r/r_m} - \left[
C_6\left(\frac{r_m}{r} \right)^6 + C_8\left(\frac{r_m}{r}
\right)^8 + C_{10}\left(\frac{r_m}{r} \right)^{10}\right] F(r)
\right\},
\end{equation}
where
\begin{eqnarray}
F(r) & = & \left\{
\begin{tabular}{lll}
$e^{-(\frac{Dr_m}{r} - 1)^2}$
&;&$\frac{r}{r_m} \leq D$\\
1 &;&$\frac{r}{r_m} > D$,
\end{tabular}
\right.
\end{eqnarray}
and
\begin{eqnarray}
\begin{tabular}{ccc}
$\frac{\epsilon}{k_B} = 10.8 K$,& &$A = 0.5448504 \times 10^6$,\\
$\alpha = 13.353384$,& &$r_m = 2.9673 A$,\\
$C_6 = 1.37732412$,& &$C_8 = 0.4253785$,\\
$C_{10} = 0.178100$,& &$D = 1.241314 \cdot$
\end{tabular}
\end{eqnarray}
The  realistic   Aziz Potential  agrees with the He-He scattering
experimental data which satisfies the following criteria:
\begin{itemize}
 \item It has a short-range repulsive part which
described by exponential form
 \item It has also  a long range
attractive tail includes  the multipole interactions.
\end{itemize}
A realistic potential between Helium atoms must have the criteria.
Our results for the free energy calculations of the liquid $^3He$
are given in Fig. (\ref{free}) (Bordbar and Hashemi, 2002).
%%%%%%%%%%%%%%%%%%%%%%%%%%%%%%%%%%%%%%%%%%%%%%%%%%%%%%%%%%%%%%%%%%%%%%%%

%********************************************************************
%***********************************************************************
\end{document}